\documentclass[journal]{IEEEtran}
\pdfoutput=1
\usepackage{cite}
\ifCLASSINFOpdf
\else
\fi
\usepackage[cmex10]{amsmath}
\usepackage{url}
\usepackage{graphicx,color}
\usepackage{amsmath,cases,overpic,url,overpic,cite,enumerate,amssymb}
\usepackage[TABTOPCAP]{subfigure}

\newcommand{\argmax}{\mathop{\rm argmax}\limits}
\def\Vec#1{\mbox{\boldmath$#1$}}

\def\nbh{{\rm nbh}}
\def\T{({\rm top})}
\def\B{({\rm bot})}

\begin{document}
\title{Real-Time Audio-to-Score Alignment of Music Performances
Containing Errors and Arbitrary Repeats and Skips}
\author{
\thanks{Citation information: DOI 10.1109/TASLP.2015.2507862,
IEEE/ACM Transactions on Audio, Speech, and Language Processing.}
\thanks{(c) 2015 IEEE. Personal use is permitted, but
republication/redistribution requires IEEE permission. See
\protect\url{http://www.ieee.org/publications_standards/publications/rights/index.html}
for more information.
}
Tomohiko~Nakamura,~\IEEEmembership{Student~Member,~IEEE,}
\thanks{T.~Nakamura is
with the Department of Information Physics and
Computing, Graduate School of Information Science and Technology,
the University of Tokyo, Tokyo 113-8656, Japan
(Tomohiko\_Nakamura$@$ipc.i.u-tokyo.ac.jp).}
Eita~Nakamura,~\IEEEmembership{Member,~IEEE,}
\thanks{E.~Nakamura is with the Graduate School of Informatics, Kyoto
University, Kyoto 606-8501, Japan
(enakamura$@$am.kuis.kyoto-u.ac.jp).}
~Shigeki~Sagayama,~\IEEEmembership{Member,~IEEE.}
\thanks{S.~Sagayama is a Professor Emeritus of University of Tokyo,
Tokyo, 113-8656, Japan and currently with the School of
Interdisciplinary Mathematical Sciences, Meiji University,
Tokyo 164-8525, Japan (sagayama$@$meiji.ac.jp).
}
}
\markboth{IEEE/ACM TRANSACTIONS ON AUDIO, SPEECH, AND LANGUAGE
PROCESSING, VOL. XX, No. YY, 2015}%
{Nakamura \MakeLowercase{\textit{et al.}}: Real-Time Audio-to-Score
Alignment of Music Performances Containing Errors and Arbitrary Repeats
and Skips}

\maketitle
\begin{abstract}
 This paper discusses real-time alignment of audio signals of music
 performance to the corresponding score (a.k.a. score following) which
 can handle tempo changes, errors and arbitrary repeats and/or skips
 (repeats/skips) in performances.
 This type of score following is particularly useful in automatic
 accompaniment for practices and rehearsals, where errors and
 repeats/skips are often made.
 Simple extensions of the algorithms
 previously proposed in the literature are not applicable in these
 situations for scores of practical length due to the problem of large
 computational complexity. To cope with this problem, we present two
 hidden Markov models of monophonic performance with errors and
 arbitrary repeats/skips, and derive efficient score-following
 algorithms with an assumption that the prior probability distributions
 of score positions before and after repeats/skips are independent from
 each other. We confirmed real-time operation of the algorithms with
 music scores of practical length (around $10000$ notes) on a modern
 laptop and their tracking ability to the input performance within $0.7$ s
 on average after repeats/skips in clarinet performance data. Further
 improvements and extension for polyphonic signals are also discussed.
\end{abstract}

\section{Introduction}
\label{sec:intro}
Real-time alignment of an audio signal of a music performance to
a given score,
also known as score following, has been
gathering attention since its first appearance in 1984
\cite{Dannenberg1984,Vercoe1984}.
Score following is a basic technique for real-time musical
applications such as automatic accompaniment, automatic score
page-turning \cite{Arzt2008} and automatic captioning to music videos.
The technique is particularly essential for automatic accompaniment,
which synchronizes an accompaniment to a performer on the fly,
referring to performance and accompaniment scores.
Automatic accompaniment enables live performance of ensemble music by
one or a few performers.
Many studies of score following have been carried out (see
\cite{Orio2003} for a review and
\cite{Pardo2005,Cont2010,Joder2010,Duan2011,Otsuka2011,Joder2011,Montecchio2011,Nakamura2013SMC08,ENakamura2014}
for recent progress).

Automatic accompaniment is particularly useful for practices, rehearsals
and personal enjoyment of ensemble music.
In these situations, performers often make errors.
Moreover, performers may want to start playing from the middle of a
score and generally make repeats and/or skips (repeats/skips).
Since errors and repeats/skips are hard to predict, a
score-following algorithm capable of handling arbitrary errors and
repeats/skips is necessary to realize an automatic accompaniment system
effective in those situations.
Our aim is to develop such an algorithm.

Treatment of errors in score following is discussed in some studies
\cite{Orio2003,Schwarz2004,Pardo2005,ENakamura2014}.
However, a detailed discussion
and a systematic evaluation of the effectiveness of the methods for
audio score following have not been given in the literature.
Score-following algorithms that can follow repeats/skips have been
proposed in \cite{Pardo2005,Oshima2005,Montecchio2011}.
The targets of these algorithms are predetermined
repeats/skips from and to specific score positions,
and treatment of arbitrary repeats/skips is not discussed nor
guaranteed.
In fact, as we will show in this paper, simple extensions of these
algorithms have the problem of large computational cost and cannot work
in real time for long scores of practical length.
Unless the problem is solved, score-following systems can only work with
limited scores with very short length or we must give up following
arbitrary repeats/skips as most of the current systems do, both of which
sacrifice the vast potential application of score following.
Therefore, it is essential to reduce the computational complexity
to follow arbitrary repeats/skips.

The authors have presented a new type of hidden Markov model
(HMM) that describes musical instrument digital interface
(MIDI) performances with errors and arbitrary repeats/skips,
and derived a computationally efficient algorithm for the HMM
\cite{ENakamura2014}.
It reduces the computational complexity with an assumption
to simplify a probability distribution of score positions before and
after repeats/skips.
While a similar model would be applicable to the audio case, further
discussions are required since audio inputs (frame-wise
discrete in time and continuous in features) significantly differ with
MIDI inputs (continuous in time and discrete in pitches) in nature.

The main contribution of this paper is to present real-time algorithms
that can follow monophonic audio performances containing arbitrary
repeats/skips and errors.
Although monophonic score following has been addressed since
\cite{Dannenberg1984,Vercoe1984},
arbitrary repeats/skips have never been discussed despite the
practical importance of their treatment as the above mentioned.
Because polyphonic score following is still an active field of research
and the extension of the present method for polyphonic performances
requires many additional issues discussed in
Sec.~\ref{sec:Discussions},
we confine ourselves to monophonic performances.

We develop a model of music performances containing
errors and arbitrary repeats/skips with an HMM.
We first discuss how various types of errors can be incorporated into
the model
(Sec.~\ref{sec:PerformanceHMM}).
Next, we extend the model to incorporate arbitrary repeats/skips.
In order to solve the problem of large computational cost for following
arbitrary repeats/skips,
two HMMs with refined topologies are presented.
We derive efficient score-following algorithms with
reduced computational complexity based on both HMMs
(Sec.~\ref{sec:FastAlgorithms}).
We demonstrate that both algorithms can work in real
time with scores of practical length on a modern laptop computer
and are effective in following performances with errors and
arbitrary repeats/skips through evaluations using
clarinet performances during practice
(Sec.~\ref{sec:Evaluation}).
We discuss possible improvements and extensions of the proposed
algorithms for polyphonic inputs (Sec.~\ref{sec:Discussions}).
Part of this study (Sec.~\ref{sec:FastAlgorithms} and a part of
Sec.~\ref{sec:Evaluation}) was reported in our previous conference paper
\cite{Nakamura2013SMC08}.

\section{Score Following for Performances with Errors}
\label{sec:PerformanceHMM}

\subsection{Variety in Audio Performance and Statistical Approach}
\label{sec:indeterminacy}
Score following is generally challenging since audio signals of
music performances widely vary even if the same score is used.
Four typical sources of variety in monophonic audio performance
are listed below.
\begin{enumerate}[(a)]
  \item {\bf Acoustic variations:}
	Spectral features of audio performances depend on musical instruments
	and are not stationary.
	In addition, audio performances usually include noise caused by
	the surrounding environment and musical instruments
	(e.g. resonance, background noise, breath noise and other acoustics).
  \item {\bf Temporal fluctuations:}
	The tempo of the performance and onset times and durations of
	performed notes deviate from
	those indicated in scores due to performer's skills, physical
	limitations of musical instruments and musical expressions.
	For example, performances during practice are often rendered in
	slow tempo to avoid errors.
  \item {\bf Performance errors:}
	Performers may make errors due to lack of performance skills
	or mis-readings of the score.
	Errors are categorized into pitch errors (substitution
	errors), dropping notes (deletion errors), adding extra notes
	(insertion errors) \cite{Dannenberg1984}.
	Besides, performers may make pauses between notes,
	for example,
	to turn a page of the score and to check the next note.
  \item {\bf Repeats/skips:}
	Performers may repeat and/or skip phrases in particular during
	practice. Furthermore, the performers generally add or
	delete a repeated section.
 \end{enumerate}
 These four sources of variety in
 monophonic audio performance
 make score following
 difficult and motivate us to study it.
 In particular, it is essential to adapt automatic accompaniment
 systems
 to the variety in order to keep synchronization to live performances.
 Although it is out of the scope of this paper, there are other sources
 of variety in music performance such as ornaments
 \cite{Cont2010,ENakamura2014,ENakamura2014ICMC,ENakamura2014arXiv}
 and improvisation
 \cite{Fremerey2010Handling,Duan2011Aligning}.

 Recent score-following systems commonly use probabilistic models such as
 HMM to capture the variety of audio performances, and their
 effectiveness has been well confirmed \cite{Orio2003}
 (and references in the Introduction).
 They are particularly advantageous to capture continuous variations of
 audio features and to handle errors which are hard to predict.
 Therefore, we take the statistical approach in this study.

\subsection{Performance HMM}
\label{sec:HMMDescript}
\begin{figure}[t]
\centering
 \includegraphics[width=\columnwidth,clip]{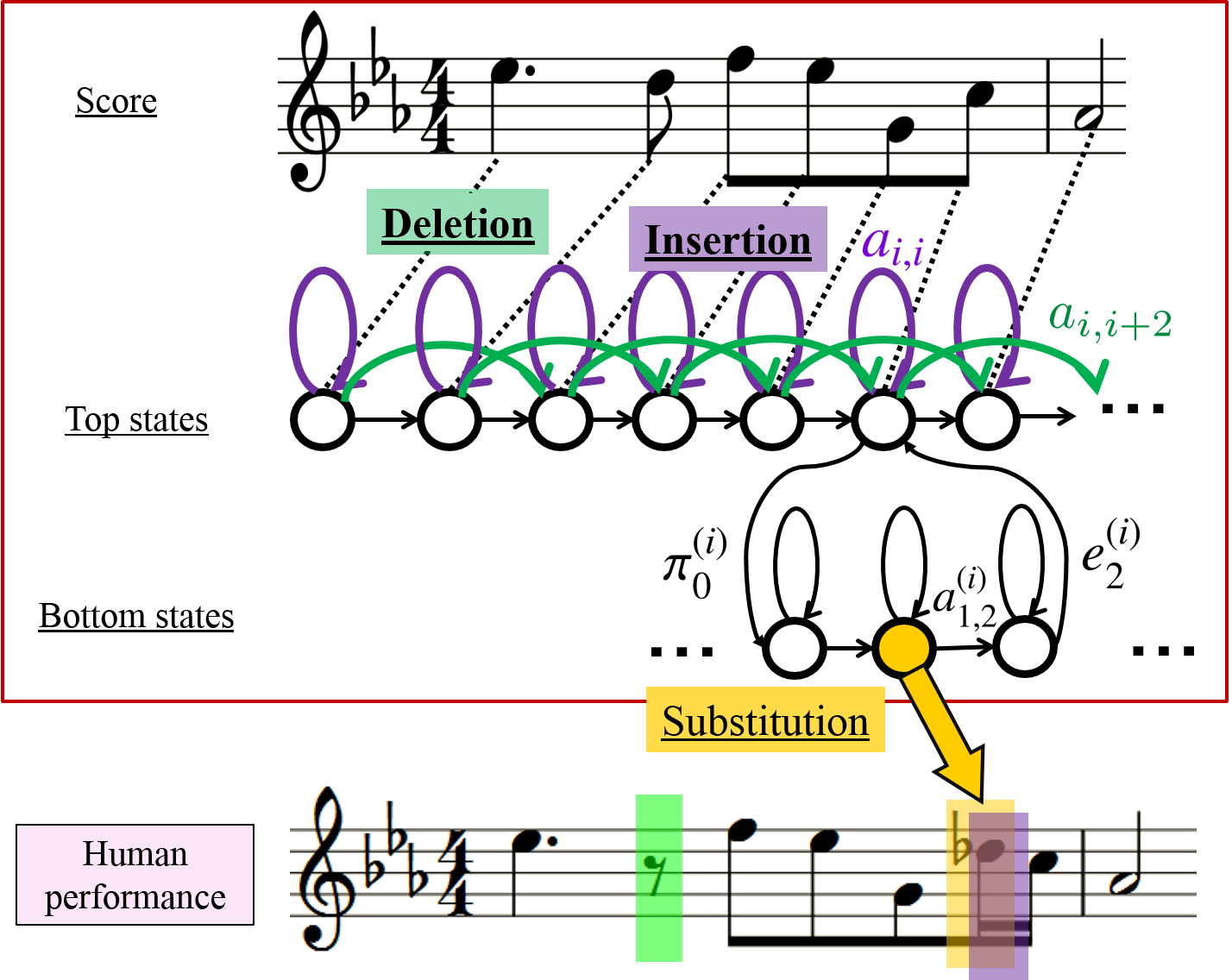}
 \caption{A hierarchical hidden Markov model
 with two levels that describes a music performance with deletion,
 insertion and substitution errors. See text.}
\label{fig:scorefollower}
\end{figure}

We represent the performance score with $N$ musical events,
each of which is a note or a rest.
A performer reads the score from event to event
and keeps making a sound corresponding to an event.
This process of performance can be modeled with a hierarchical HMM with two
levels \cite{Orio2001,Cont2006}, which we call the performance HMM.
The top level describes the progression of performed events, and the
bottom level expresses temporal structure of the audio signal in a
performed event.

Events correspond to states (top states) of the top-level HMM (top HMM),
and the performance is described as transitions between the top states.
Let $z_t^{\T}=0,\cdots,N-1$ denote the random variable describing the
top state at the $t$th frame ($t=0,\cdots,T-1$), and let $i$ and
$j$ label a top state.
The top HMM is parameterized by state transition probabilities $a_{j,i}$
and initial probabilities $\pi_{i}$:
\begin{align}
 a_{j,i}&:=P(z_{t}^{\T}=i|z_{t-1}^{\T}=j),\\
 \pi_{i}&:=P(z_{0}^{\T}=i),
\end{align}
which satisfy $\sum_{i=0}^{N-1}\pi_{i}=1$ and $\sum_{i=0}^{N-1}a_{j,i}=1$
for all $j$.

Each top state is itself an HMM (bottom HMM), whose states
(bottom states) correspond to subevents in an event, for example,
sustain of an instrumental sound, pauses between notes, etc.
Let $L$ denote the number of bottom states in the top state,
$z_t^{\B}=0,\cdots,L-1$ denote the random variable describing the
bottom state at the $t$th frame, and let $l$ and $l'$ label a
bottom state.
The state transitions of the bottom HMM are characterized by
three kinds of probabilities.
The initial probability $\pi_{l}^{(i)}$ describes the probability of a
transition to bottom state $l$ when top state $i$ is entered,
the exiting probability $e^{(i)}_{l}$ describes the probability of
exiting top state $i$ from bottom state $l$, and the transition
probability
$a_{l',l}^{(i)}:=P(z_{t}^{\B}=l|z_{t-1}^{\B}=l')$ represents the
transition from bottom state $l'$ to bottom state $l$ in top state $i$.
These probabilities satisfy
$\sum_{l'=0}^{L-1}\pi_{l'}^{(i)}=1$ and
$\sum_{l'=0}^{L-1}a_{l,l'}^{(i)}+e^{(i)}_{l}=1$
for all $l$ and $i$.
Thus, the performance is modeled as a sequence of $T$ pairs of random variables
$\{(z_{t}^{\T},z_{t}^{\B})\}_{t=0}^{T-1}$ (Fig.~\ref{fig:scorefollower}).
For example, if the pair $z_{t}:=(z_{t}^{\T},z_{t}^{\B})$ equals to $(i,l)$,
the score position at frame $t$ is at bottom state $l$ of top state $i$.
 
Observed audio features are described as being stochastically
generated from a bottom state.
Given an audio feature ${\mathbf
y}_{t}:=[y_{t,0},y_{t,1},\cdots,y_{t,D-1}]^{\top}$ at frame $t$ as a
$D$-dimensional real vector,
the emission probability of state $(i,l)$ is defined as
\begin{align}
 b_{l}^{(i)}({\mathbf y}_{t})&:=P({\mathbf y}_{t}|z_{t}=(i,l)).
\end{align}

\subsection{Emission Probability and Substitution Error}
\label{sec:model_subst}
From here to Sec.~\ref{sec:model_pauseins}, we consider the performance
HMM with $L=1$ for simplicity, but the case for $L>1$ can be treated
similarly.
To extract pitch information from the input signal, we need a suitable
feature representation.
In the comparison of some audio features in \cite{Joder2010,Joder2013},
the magnitude of a constant-Q transform (CQT) \cite{Brown1992} with a
quality factor set to one semitone yielded the best result of score
following for monophonic audio input.
Furthermore, normalizing magnitudes of CQTs such that
$\sum_{d=0}^{D-1}y_{d}=1$ makes them insusceptible to dynamic
variations.
Although one may think that the normalization makes it difficult to
discriminate pauses from notes, the difference in spectral shape
between pauses and notes can help the discrimination:
The CQT of a pitched sound have clear peaks at its fundamental
frequency and harmonics, whereas the CQTs at pauses are
relatively flat.
We use normalized magnitudes of CQTs (normalized CQTs) as audio
features.

Let $k$ be the pitch index and ${\cal K}$ be the set of possible pitches.
For convenience, we indicate the pitches A0 to C8 in the range of a
standard piano as $k=21$ to $k=108$ and silence as $k=-1$, and
${\cal K}=\{21,22,\cdots,108\}\cup\{-1\}$.
We assume that normalized CQTs corresponding to pitch $k$
follow a $D$-dimensional normal distribution with mean
$\Vec{\mu}_{k}$ and covariance matrix $\Sigma_{k}$,
denoted by ${\cal N}({\mathbf y}_{t}|\Vec{\mu}_{k},\Sigma_{k})$.
The emission probability $b_{0}^{(i)}(\mathbf{y}_t)$
of bottom state $0$ of top state $i$
is given as
\begin{equation}
 b_{0}^{(i)}(\mathbf{y}_t)=\sum_{k\in{\mathcal{K}}}w_{k,0}^{(i)}{\cal
  N}(\mathbf{y}_t|\Vec{\mu}_{k},\Sigma_{k}).
  \label{eq:def_emsprob}
\end{equation}
Here $w_{k,0}^{(i)}\in[0,1]$ is a mixture weight of pitch $k$
of bottom state $0$ of top state $i$,
which satisfies $\sum_{k\in\mathcal{K}}w_{k,0}^{(i)}=1$ for all $i$.

When substitution errors are not made, $w_{k,0}^{(i)}=0$ unless
$k=p_{i}$, where $p_{i}\in{\cal K}$ denotes the pitch of event $i$
($p_{i}=-1$ for a rest).
On the other hand, to describe a performance with substitution errors,
we have small positive values of $w_{k,0}^{(i)}$ for $k\neq p_{i}$
since a substitution error is represented by an emission of
an audio feature with an incorrect pitch.

\subsection{Transition Probability and Deletion and Insertion Errors}
\label{sec:model_delins}
Transition probabilities in the top level $a_{j,i}$ represent the
frequency of the transitions between the events.
If performances do not contain insertion and deletion errors,
$a_{j,i}=0$ unless $i=j+1$.
We can express an insertion error and a deletion error
with a self transition and a transition to the second next top state,
which correspond to $a_{j,j}$ and $a_{j,j+2}$.

The self-transition probability $a_{0,0}^{(i)}$ of
bottom state $0$ of top state $i$ describes the expected duration of the
corresponding event $d_{i}$, which is computed as a product of the note
value of the event and the score-notated tempo:
\begin{equation}
 d_{i}=\sum_{k=1}^{\infty}k(a^{(i)}_{0,0})^{k-1}(1-a^{(i)}_{0,0})
  =\dfrac{1}{1-a^{(i)}_{0,0}}.
\label{eq:def_ai00}
\end{equation}
If $d_{i}$ is shorter than a processing time interval,
we put $a_{0,0}^{(i)}=0$.
This probabilistic representation of the event duration
describes the temporal fluctuations of music performance.

\subsection{Pauses between Notes}
\label{sec:model_pauseins}
\begin{figure}[t]
\centering
\includegraphics[width=0.8\columnwidth,clip]{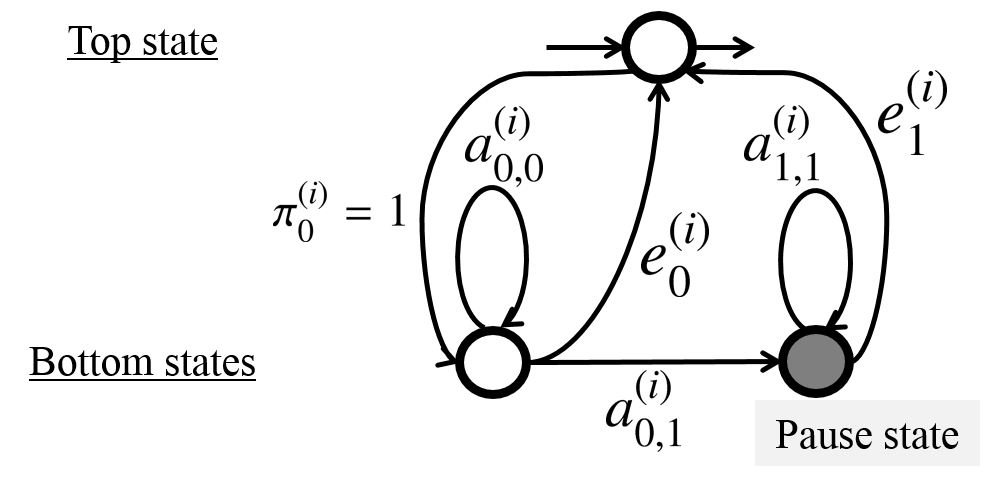}
 \caption{A pause between notes is described with the pause state
 (gray disk) which emits audio features corresponding to silence.}
\label{fig:lowerlayer_topology}
\end{figure}

Pauses between notes can be introduced into the performance HMM by
adding an extra bottom state with index $1$,
which we call a pause state (Fig.~\ref{fig:lowerlayer_topology}).
The occurrence of the pause is expressed as a transition to the pause
state, which corresponds to $a_{0,1}^{(i)}$.
The duration of the extra pause is represented by the self-transition
probability of the pause state $a_{1,1}^{(i)}$,
which can be set similarly to Eq.~\eqref{eq:def_ai00}.
We put $a_{1,0}^{(i)}=0$ and $\pi_{1}^{(i)}=0$ for all $i$.
We assume that $b_{1}^{(i)}(\mathbf{y}_{t})={\cal
  N}(\mathbf{y}_t|\Vec{\mu}_{-1},\Sigma_{-1})$.

\subsection{Estimation of Score Positions}
\label{sec:forward}
For the convenience of estimating score positions,
we convert the performance HMM into an equivalent standard HMM.
Its state corresponds to a bottom state of the
performance HMM and is labeled with $(i,l)$.
The standard HMM is parameterized by emission probabilities
$\tilde{b}_{(i,l)}(\mathbf{y}_{t})$, initial probabilities
$\tilde{\pi}_{(i,l)}$, and transition probabilities
$\tilde{a}_{(j,l'),(i,l)}$, defined by
$\tilde{b}_{(i,l)}(\mathbf{y}_t):=b_{l}^{(i)}(\mathbf{y}_t)$,
$\tilde{\pi}_{(i,l)}:=\pi_{i}\pi^{(i)}_{l}$, and
\begin{align}
\tilde{a}_{(j,l'),(i,l)}&:=
\begin{cases}
 a_{l',l}^{(i)}+e_{l'}^{(i)}a_{i,i}\pi_{l}^{(i)} & (i=j) \\
 e_{l'}^{(j)}a_{j,i}\pi_{l}^{(i)} & (i\neq j)
\end{cases},
\label{eq:def_tildea}
\end{align}

Given observed normalized CQTs up to the
$t$th frame $\mathbf{y}_{0:t}=\{\mathbf{y}_{\tau}\}_{\tau=0}^{t}$,
the score position at frame $t$ is estimated with the standard HMM by solving
\begin{align}
 \argmax_{z_t}
 P(z_t|\mathbf{y}_{0:t})
 &=
 \argmax_{z_t}
 P(\mathbf{y}_{0:t},z_{t}),
\label{eq:problem_formulate}
\end{align}
where
\begin{align}
 P(\mathbf{y}_{0:t},z_{t})
 &=\sum_{z_{0:t-1}}
 \Big(\prod_{\tau=1}^{t}\tilde{b}_{z_{\tau}}(\mathbf{y}_{\tau})
 \tilde{a}_{z_{\tau-1},z_{\tau}}\Big)
 \tilde{b}_{z_{0}}(\mathbf{y}_{0})\tilde{\pi}_{z_{0}}.
 \label{eq:arranged}
\end{align}
Here $z_{0:t-1}$ denotes $\{z_{\tau}\}_{\tau=0}^{t-1}$.
Eq.~\eqref{eq:problem_formulate} is derived from the Bayes' theorem.

This maximization problem can be solved efficiently with the forward
algorithm.
It computes the forward variable
$\alpha_{t,z_{t}}:=P(\mathbf{y}_{0:t},z_{t})$
in a recursive manner:
\begin{align}
\alpha_{t,(i,l)}=
\begin{cases}
\displaystyle
\tilde{b}_{(i,l)}(\mathbf{y}_{t})
\sum_{\substack{j=0,\cdots,N-1\\ l'=0,\cdots,L-1}}
\alpha_{t-1,(j,l')}
\tilde{a}_{(j,l'),(i,l)}
& (t\geq 1), \\
\tilde{b}_{(i,l)}(\mathbf{y}_0)\tilde{\pi}_{(i,l)}
&(t=0).
\end{cases}
\label{eq:ForwardAlgorithm_layered}
\end{align}
Since $\tilde{a}_{(j,l'),(i,l)}=0$ unless $0\leq i-j\leq 2$,
the complexity of computing $\alpha_{t,(i,l)}$ is
of ${\cal O}(LN)$ at each time step.

\section{Incorporating Arbitrary Repeats/Skips and Fast Score-Following Algorithms}
\label{sec:FastAlgorithms}
\subsection{Incorporating Arbitrary Repeats/Skips and Computational
  Complexity for Inference}
\label{sec:comp_comp}
So far, the top HMM is left-to-right and its states are connected only
to their neighboring states.
However, all top states must be connected to describe arbitrary
repeats/skips, i.e. $a_{j,i}>0$ for all $j$ and $i$.
The model is a generalization of the performance models in previous
studies \cite{Pardo2005,Oshima2005,Montecchio2011}.

Assuming $L=1$ for simplicity
and dropping the subscripts $l,l'$ from the parameters of the
standard HMM and the forward variables as
$\tilde{a}_{j,i}:=\tilde{a}_{(j,0),(i,0)}$,
$\tilde{b}_{i}(\mathbf{y}_{t}):=\tilde{b}_{(i,0)}(\mathbf{y}_{t})$,
$\tilde{\pi}_{i}:=\tilde{\pi}_{(i,0)}$
and $\alpha_{t,i}:=\alpha_{t,(i,0)}$,
Eq.~\eqref{eq:ForwardAlgorithm_layered} can be rewritten as
\begin{align}
\alpha_{t,i}=
\begin{cases}
\displaystyle
\tilde{b}_{i}(\mathbf{y}_{t})
\sum_{j=0}^{N-1}
\alpha_{t-1,j}
\tilde{a}_{j,i}
&(t\geq 1), \\
\tilde{b}_{i}(\mathbf{y}_0)\tilde{\pi}_{i}
&(t=0).
\end{cases}
\label{eq:ForwardAlgorithm}
\end{align}
Eq.~\eqref{eq:ForwardAlgorithm} for $t\geq1$ contains
a summation over $N$ states for each $i$,
and the complexity is of ${\cal O}(N^2)$.
As we will experimentally show in Sec.~\ref{exp:comp_time},
this complexity is too large to run in real time with scores of
practical length on a modern laptop.
Therefore, it is crucial to reduce the complexity.
It is noteworthy that a similar large complexity can emerge
even if only specific repeats/skips are allowed
(e.g. transitions between the first notes of bars in a score),
since the number of such specific transitions often increases
in proportion to $N$.

One may think that pruning techniques can be used to reduce the
computational complexity.
However, pruning is ineffective here since repeats/skips seldom occur,
and it is necessary to take all transitions into account.
Computing all transitions has a benefit also in
following performances without repeats/skips.
When an estimation error of score position occurs,
a score follower may fail to track the performance and become lost.
It often happens that a score follower with a pruning technique
(e.g. with a limited search window)
cannot recover from being lost.
By contrast, if a score follower searches all transitions,
it can return to find the correct score position after a while if the
performer continues the performance.

\subsection{Reduction of Computational Complexity by Factorizing
  Probabilities of Repeats/Skips}
\label{sec:outerproduct}
One method to reduce the computational complexity while computing all
transitions is to introduce some constraints on the transition
probabilities.
In \cite{ENakamura2014},
reduction of the computational complexity
is achieved with an assumption that
the probability of score positions where performers stop before
repeats/skips (stop positions) is the same regardless of where they resume
performing after repeats/skips (resumption positions).

We shall introduce this assumption to the performance HMM.
The transition probability of a repeat/skip from event $j$ to event $i$
is then written as a product of two probabilities $s_{j}$ and $r_{i}$.
$s_{j}$ is the probability of stopping at event $j$ before a
repeat/skip,
and $r_{i}$ is the probability of resuming a performance at event $i$
after a repeat/skip.
The transition probability of the top HMM is then written as
\begin{align}
a_{j,i}=a^{(\nbh)}_{j,i}+s_{j}r_{i}.
\label{eq:factorize_transprob}
\end{align}
where $a^{(\nbh)}_{j,i}$ is a band matrix satisfying $a^{(\nbh)}_{j,i}=0$
unless $0\leq i-j \leq 2$.
The parameter $a_{j,i}^{(\nbh)}$ characterizes transitions within
neighboring states and is determined according to the normalization
constraint of $a_{j,i}$,
which is written as
$1=\sum_{i}a_{j,i}=\sum_{i}a^{(\nbh)}_{j,i}+s_{j}\sum_{i}r_{i}$ for all $j$.
Without loss of generality, we can assume $\sum_{i}r_{i}=1$ and then
we have $\sum_{i}a^{(\nbh)}_{j,i}=1-s_{j}$.

Let us denote the set of neighboring states of top state
$i$ by $\nbh(i):=\{j;j=0,\cdots,N-1,\ 0\leq i-j \leq 2\}$.
The transition probability of the standard HMM $\tilde{a}_{j,i}$
for $j\notin\nbh(i)$ is written as
\begin{align}
 \tilde{a}_{j,i}=e^{(j)}_{0}s_{j}r_{i}\pi_{0}^{(i)}.
 \label{eq:outerproduct_transprob}
\end{align}
With Eqs.~\eqref{eq:outerproduct_transprob} and \eqref{eq:ForwardAlgorithm}, we have
\begin{align}
\alpha_{t,i}
=&
\tilde{b}_{i}(\mathbf{y}_t)
\Big\{
 \sum_{j\in\nbh(i)}
 \alpha_{t-1,j}
 \tilde{a}_{j,i}
\notag
\\
&
+
r_{i}\pi_{0}^{(i)}
\Big(
 \sum_{j=0}^{N-1}
 \alpha_{t-1,j}e_{0}^{(j)}s_{j}
 -
 \sum_{j\in\nbh(i)}
 \alpha_{t-1,j}e_{0}^{(j)}s_{j}
\Big)
\Big\}.
\label{eq:OuterProductHMM_Forward}
\end{align}
Since the first summation in the parentheses of the second term is independent
of $i$,
it is sufficient to calculate it once at each time step.
This term and the rest of Eq.~\eqref{eq:OuterProductHMM_Forward} are of
${\cal O}(N)$, and hence
the total computational complexity is ${\cal O}(N)$.
The space complexity is also reduced: 
The transition probability matrix in the top level is
now parameterized by $4(N-1)$ parameters ($s_{j}$, $r_{i}$ and $a^{(\nbh)}_{i,j}$).
It has $N(N-1)$ parameters originally.

This result can be generalized for the performance HMM with $L>1$.
The standard HMM has $LN$ states and updating $\alpha_{t,(i,l)}$ at each
time step is of ${\cal O}((LN)^2)$
according to Eq.~\eqref{eq:ForwardAlgorithm_layered}.
If we introduce the above assumption, the transition probability of the
standard HMM $\tilde{a}_{(j,l'),(i,l)}$ can also be divided into
a component dependent only on $i,l$ and a component dependent only on $j,l'$.
Therefore, the total computational complexity is reduced to ${\cal
O}(LN)$ (see Appendix~\ref{sec:twolayeredHMM_outerproduct} for details).
Importantly, this reduction method can be used regardless of the topology of the
bottom HMMs, and it is compatible with
the pause states and applicable to performance HMMs with
more complex structure of bottom HMMs
(e.g. \cite{Cano1999,Raphael1999,Orio2001,Cont2010}).

A similar reduction method is valid for the Viterbi algorithm and the
backward algorithm.
The method can be applied to any HMM and similar dynamic programming
techniques as well,
and it can be useful for applications other than score following,
(e.g. timbre editing of music signals \cite{Nakamura2014ICASSP05}).

\subsection{Explicit Description of Silent Breaks at Repeats/Skips}
\label{sec:break}
\begin{figure}[t]
\centering
\includegraphics[width=\columnwidth,clip]{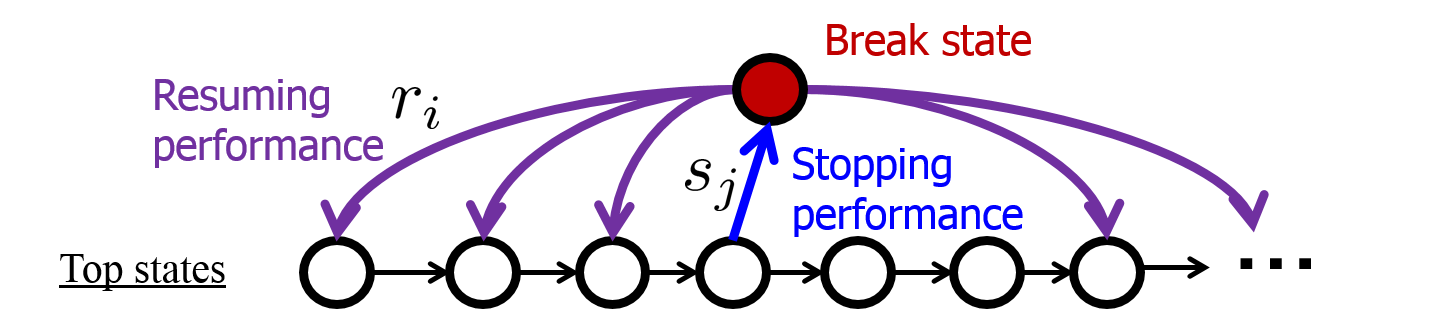}
 \caption{A repeat/skip can be described with two-step
 transitions via the break state representing silent breaks.}
\label{fig:wpause_model}
\end{figure}

We can achieve a similar reduction of the computational complexity by
using another assumption on arbitrary repeats/skips.
Performers frequently make silent breaks at repeats/skips
to get ready for resuming the performance.
In fact, $59$ of $63$ repeats/skips accompanied the breaks longer than
$500$ ms in actual performances used in
Sec.~\ref{sec:prepare_realperf}.

Let us represent the silent breaks by introducing an additional state
(the break state) as the $N$th top state.
The duration of the breaks is described with
the self-transition probability of the bottom state of the break
state $a_{0,0}^{(N)}$,
and its value is determined similarly to Eq.~\eqref{eq:def_ai00}.
Repeats/skips are represented as two-step transitions via the break
state (Fig.~\ref{fig:wpause_model}).
Stopping (resuming) a performance is expressed as
transitions to (from) the break state
whose probability is denoted by $s_{j}$ ($r_{i}$, respectively).
We note that the top states excluding the break state are
connected only to neighboring top states,
and thus $\tilde{a}_{j,i}=0$ if $j\notin \nbh(i)$
for all $i,j\neq N$.
On the other hand, the break state is connected to all top states except
itself.
We put $\tilde{a}_{N,N}=0$.
The transition probability of the standard HMM from or to the break
state is written as
\begin{align}
\tilde{a}_{j,N}=&e^{(j)}_{0}s_{j}\pi_{0}^{(N)}
\quad (j\neq N),
\\
 \tilde{a}_{N,i}=&
 \begin{cases}
  e^{(N)}_{0}r_{i}\pi_{0}^{(i)} &(i\neq N), \\
  a_{0,0}^{(N)} & (i=N)
 \end{cases}
\end{align}
where $e^{(N)}_{0}(=1-a_{0,0}^{(N)})$ and $\pi_{0}^{(N)}(=1)$ denote the
exiting probability and the initial probability of state $(N,0)$.

For this model,
Eq.~\eqref{eq:ForwardAlgorithm} for $t\geq 1$ can be written as
\begin{equation}
 \alpha_{t,i}=
\begin{cases}
\displaystyle
\tilde{b}_{i}(\mathbf{y}_{t})
\Big(
\sum_{j\in \nbh(i)}
\alpha_{t-1,j}
\tilde{a}_{j,i}
+
\alpha_{t-1,N}\tilde{a}_{N,i}
\Big)
& (i\neq N) \\
\displaystyle
\tilde{b}_{N}(\mathbf{y}_{t})
\sum_{j=0}^{N-1}
\alpha_{t-1,j}\tilde{a}_{j,N}
& (i=N).\\
\end{cases}
\label{eq:ForwardAlgorithm_wsilent}
\end{equation}
We see that updating $\alpha_{t,i}$
involves summation of at most four terms for each $i\neq N$
and $N$ terms for $i=N$.
The total complexity is thus ${\cal O}(N)$ for each time step.
This reduction method can also be extended to the case of $L>1$
(see Appendix~\ref{sec:twolayeredHMM_proposed}).

It is noteworthy that the performance HMM with the break state
is related to the performance HMM presented in
Sec.~\ref{sec:outerproduct}.
If we assume that transitions go through the break state in no
time, the two-step transition from top state $j$ to top
state $i$ via the break state is reduced to the
direct transition from top state $j$ to
top state $i$,
and its probability is written as a product of $s_{j}$ and $r_{i}$.
In other words, the difference between these models is whether
breaks are explicitly described.
Since it is difficult to quantify its effect on the performance of score
following analytically, we will evaluate the effect through an
experiment in Sec.~\ref{sec:scofo_perf_result}.

\section{Experimental Evaluation of the Proposed Score-Following Algorithms}
\label{sec:Evaluation}
\subsection{Processing Time}
\label{exp:comp_time}
\begin{figure}[t]
\centering
 \includegraphics[width=0.9\columnwidth,clip]{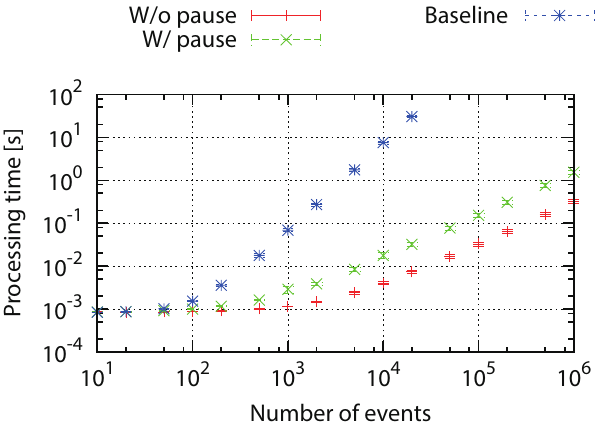}
 \caption{Average processing times with standard errors with respect to
 the number of events.
 ``W/ pause'' and ``W/o pause'' represent
 the {\em break algorithm} with and without the pause states, respectively.
 ``Baseline'' represents a simple extension of the algorithms proposed
 in previous studies \cite{Pardo2005,Oshima2005,Montecchio2011}.}
 \label{fig:compare_comptime}
\end{figure}

We measured processing times in order to evaluate the reduction of the
computational complexity with the proposed algorithms.
The processing time depends on the number of events $N$ and virtually
not on other score content and signal content.
We used synthetic scores with $10$ to $10^6$ events\footnote{Practical
scores contain ${\cal O}(10^3)$ to ${\cal O}(10^4)$ notes. For
instance, there are around $2200$ events in the
clarinet part of the first movement in the Mozart's Clarinet Quintet.}
and a random signal of two seconds length
with a sampling rate of $16$ kHz as an audio input.
Normalized CQTs were computed with a frame length of $128$ ms and a
hopsize of $20$ ms.
Their center frequencies ranged from $55$ to $7040$ Hz at a semitone
interval, and the quality factor was set to $16$, which approximately
corresponds to one semitone.
Algorithms were implemented in C++ on a computer with $3.30$ GHz
CPU (Intel(R) Core(TM) i3-2120 CPU) and $8$ GB memory running Debian.

Processing times averaged over $100$ frames with standard errors
are shown in Fig.~\ref{fig:compare_comptime} for the algorithms proposed
in Sec.~\ref{sec:break} ({\em break algorithm}) with and without the
pause states ($L=2$ and $L=1$) and the algorithm that calculates
$\alpha_{t,i}$ according to Eq.~\eqref{eq:ForwardAlgorithm}
(baseline algorithm).
(The results for the algorithm proposed in Sec.~\ref{sec:outerproduct}
({\em no-break algorithm}) did not significantly differ with the results for the {\em break algorithms}.)
It can be confirmed that the average processing times increased
asymptotically in proportion to $N^2$ ($N$) with the baseline algorithm
(the {\em break algorithms}, respectively).
The result shows that the proposed algorithms
significantly suppress the increase of processing times.
The processing times for $N\geq 1000$ were larger than the hopsize with
the baseline algorithm,
and the algorithm can work in real time
with scores with only up to ${\cal O}(10^2)$ events,
which is the size of short music pieces.
By contrast, the average processing times were below the hopsize for
$N\leq 10000$ ($N\leq 50000$) with the {\em break algorithm} with
(without, respectively) the pause states.
Therefore, the proposed algorithms with and without the pause states can
work in real time with scores with up to ${\cal O}(10^3)$ events and
${\cal O}(10^4)$ events, respectively.
Note that processing times
depend on the computing power, but their relative values remain almost
the same and the proposed algorithms are always effective in reducing
 the computational complexity.

\subsection{Score-Following Accuracy for Performances with Errors}
\label{sec:exp_Bach10}

\subsubsection{Data Preparation}
\label{sec:error_simulate_Bach10}
To evaluate the score-following accuracy for performances with errors,
we conducted an experiment using the Bach10 dataset
\cite{Duan2011Soundprism}.
It consists of audio recordings of ten four-part chorales by J.~S. Bach.
The soprano, alto, tenor and bass parts of each piece were
separately recorded and performed by the violin,
clarinet, saxophone and bassoon, respectively.
Their durations ranged from $25$ to $41$ seconds.

Since the performances did not contain errors,
we simulated errors by randomly inserting, dropping and
substituting notes in each score, which correspond to deletion, insertion
and substitution errors in the performance, respectively.
Their probability values were obtained from the MIDI piano performances
during practice in \cite{ENakamura2014}:
$0.0034$ for deletion errors and $0.0245$ for insertion errors.
For simplicity, substitution errors were restricted to three types
typical in clarinet performances, namely errors in semitone,
whole-tone and perfect 12th.
The first two errors are often caused by fingering errors and
mis-readings of the score,
and the last error is caused by overblowing on a clarinet.
The probability values of the three pitch errors were $0.0145$, $0.0224$ and
$0.0047$ in the simulation, where the probability of the perfect 12th
pitch error was substituted by that of the octave pitch errors obtained in
\cite{ENakamura2014}.

\subsubsection{Experimental Conditions}
\label{sec:expcond_Bach10}
We conducted a preliminary experiment
and set the parameter for performance errors as follows:
$a_{i,i+2}=1.0\times10^{-50}$ for deletion errors,
$a_{i,i}=0$ for insertion errors,
and $a_{1,1}^{(i)}=0.999$ and $a_{0,1}^{(i)}=1.0\times10^{-100}$ for
pauses between notes.
Although the mixture weight $w_{k,0}^{(i)}$ can be learned from audio
signals at each $k$ and $i$ in principle, it is difficult to obtain them
independently for the lack of enormous performance data.
To reduce the number of parameters, we considered only the most
important three substitution errors described in the previous section.
The mixture weights $w_{k,0}^{(i)}$ for the errors were designed
in proportion to their frequencies used in the simulation:
\begin{align}
 w_{k,0}^{(i)}=
 \begin{cases}
  1-C & (k=p_{i})\\
  C\times 0.175 & 
  (k=p_{i}\pm 1)\\
  C\times 0.270 & 
  (k=p_{i}\pm 2)\\
  C\times 0.055 & 
  (k=p_{i}\pm 19)\\
  0 & (\text{otherwise})
  \label{eq:design_C}
 \end{cases}
\end{align}
for all $p_{i}\neq -1$, where $C$ is the probability of pitch errors.
The value of $C$ was optimized in a preliminary experiment and
we set $C=1.0\times 10^{-50}$.
For $p_{i}=-1$, we put $w_{k,0}^{(i)}=0$ unless $k=-1$.
The probabilities of stopping and resuming a performance
$s_{j},r_{i}$ were set uniformly in $i,j$:
$s_{0}=s_{1}=\cdots=s_{N-1}=1.0\times10^{-x}$ for some positive $x$
and $r_{0}=r_{1}=\cdots=r_{N-1}=1/N$.
Since the value of $a_{0,0}^{(N)}$ did not significantly change the
result in a preliminary experiment, we fixed $a_{0,0}^{(N)}=0.996$.

The accuracy of score following generally depends on
the parameters of the emission probabilities.
It has been reported that learning them from audio performances improves
the accuracy \cite{Joder2011,Joder2013},
and thus we learned the parameters $\Vec{\mu}_{k}$ and $\Sigma_{k}$
from audio signals.
The parameters can be learned from every musical instrument if
necessary data is available and we can form a detailed model for a specific
instrument.
Alternatively, we can use a set of data consisting of several musical
instruments to form a ``general model'' that can be applied for a wider
class of instruments.
Such a learning method is applicable for any instruments in principle,
and it can be even more effective for musical instruments with complex
signals, for which physical modeling or manual spectrum-template
construction is more difficult.
In general, there is a tradeoff between the generalization capability
and the adaptation ability.
Here, we learned the parameters with performance data of several musical
instruments and used them to measure the accuracy of score following.

The learning data consisted of performances played by the
violin and clarinet in RWC musical instrument database \cite{Goto2004}.
To reduce overfitting, we assumed that $\Sigma_{k}$ is diagonal
and introduced a lower bound, or a flooring value $F$,
on the diagonal elements of $\Sigma_{k}$.
The introduction of $F$ is called the flooring method
and generally used for speech recognition (e.g. see \cite{HTK}).
We conducted a preliminary experiment and found the optimal
$F=1.0\times10^{-4}$.
The initial probabilities were set as $\pi_{i}=0$ for $i\neq 0$ and
$\pi_{0}=1$.

We compared the proposed algorithms with {\em Antescofo} \cite{Cont2010},
which is one of the most known score-following systems applied
to various musical pieces and used in the most severe artistic
situations.
Antescofo was not developed to cope with repeats/skips in monophonic
performances, and is without special treatments for repeats/skips.
It had the best accuracy in the music information retrieval
evaluation exchange (MIREX 2006) \cite{MIREX}, which is the most famous
evaluation contest in this field.
Since Antescofo ended score following when the last note in the
score was estimated,
estimated score positions were assumed to be the last note
from the time when Antescofo ended score following.	
The overall accuracy of score following was measured by piecewise
precision rate (PPR), defined as the piecewise average rate of
onsets correctly detected within $\Delta$ ms error.
The PPR has been used with $\Delta=2000$ ms in MIREX
\cite{Cont2007,MIREX}.

\subsubsection{Results}
\begin{figure}[t]
 \centering
 \includegraphics[width=\columnwidth]{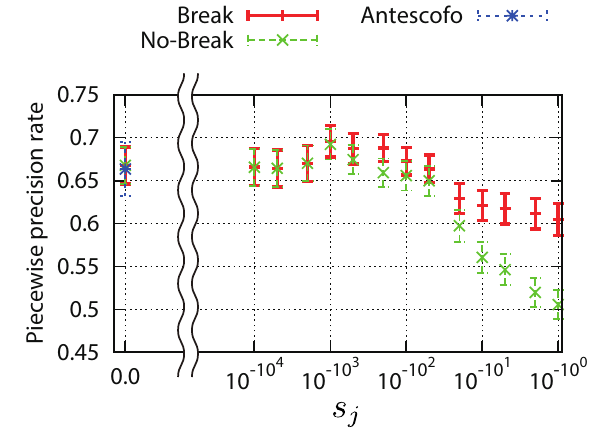}
 \caption{
 Average piecewise precision rates and standard errors with
 respect to $s_j$ for audio performances obtained by simulating errors.
 The {\em break algorithm} (``Break'') and the {\em no-break algorithm}
 (``No-Break'') without the pause states are compared to Antescofo
 \cite{Cont2010}.}
 \label{fig:evalppr_Bach10}
\end{figure}

Tab.~\ref{tab:instrumentwise_Bach10} summarizes average PPRs and
standard errors with $\Delta=300$ ms for every musical instrument.
The results for the {\em no-break algorithm} did not significantly
differ with the results for the {\em break algorithm} when $s_j=0.0$.
We found that the proposed algorithms provided similar accuracies for
the saxophone and bassoon data, which were not contained in the learning
data, compared to the clarinet and violin data.
The PPRs obtained with the proposed algorithms were similar to those
obtained with Antescofo in all data.

Fig.~\ref{fig:evalppr_Bach10} illustrates average PPRs and standard
errors with $\Delta=300$ ms.
As described in Sec.~\ref{sec:comp_comp}, computing all transitions help
that the score follower returns to recover from being lost.
The benefit can be confirmed from that the proposed algorithms with
$s_{j}=1.0\times 10^{-1000}$ provided around $0.05$ higher accuracy than
Antescofo, which searches only local transitions.
On the other hand, $s_{j}$s larger than $1.0\times 10^{-500}$ caused the
frequent overdetection of repeats/skips and the accuracy became lower
than $s_{j}=0$.
A similar tendency was observed in PPR with $\Delta=500$ and $2000$ ms.

Large values of $s_{j}$ deteriorated the score-following accuracy of the
present algorithms as shown in Fig.~\ref{fig:evalppr_Bach10}.
This is because the larger $s_{j}$, the more frequently the
algorithms may misestimate insertion/deletion/substitution errors as
repeats/skips.
We indeed confirmed that the number of misdetected repeats/skips
increased with larger $s_{j}$.

There was around $0.1$ difference in PPR between the algorithms when
$s_{j}$ is large.
We found that the total number of misdetected repeats/skips by the
{\em no-break algorithm} was around $1.2$ times larger than that of the
{\em break algorithm} for $s_{j}\geq 10^{-10}$.
Since the {\em break algorithm}
assumes that repeats/skips always
accompany breaks and simulated errors did not accompany pauses,
the results suggest that the explicit description of the breaks
reduced misestimations of the errors as repeats/skips.

 \begin{table}[t]
  \centering
 \caption{Average piecewise precision rates and standard errors for
 violin, clarinet, saxophone and bassoon performances with errors.
 ``Proposed ($s_j=0$)'' (``Antescofo'') denotes the \emph{break
 algorithm} with $s_{j}=0$ (Antescofo \cite{Cont2010}, respectively).
 }
 \begin{tabular}{c|ccc}
  Musical instrument &
  Proposed ($s_j=0$) &
  Antescofo \\ \hline
  Violin & $0.72\pm0.03$ & $0.66\pm 0.06$ \\
  Clarinet & $0.61\pm0.05$ & $0.57\pm0.08$ \\
  Saxophone & $0.63\pm0.06$ & $0.64\pm0.06$ \\
  Bassoon & $0.76\pm0.04$ & $0.79\pm0.04$ \\
 \end{tabular}
 \label{tab:instrumentwise_Bach10}
 \end{table}

\subsection{Score-Following Accuracy for Performances with Errors and Repeats/Skips}
\label{sec:scofo_perf}

\subsubsection{Performance Data During Practice}
\label{sec:prepare_realperf}
\begin{table}[t]
 \centering
 \caption{The number of errors and repeats/skips in the used clarinet
 performances.}
 {\tabcolsep=3pt
  \begin{tabular}[t]{c|ccccc}
  &\begin{tabular}{c}
   Pauses\\
   between notes
  \end{tabular}
  &\begin{tabular}{c}
    Deletion\\
    error
   \end{tabular}
  &\begin{tabular}{c}
    Insertion\\
    error
  \end{tabular}
  &\begin{tabular}{c}
    Substitution\\
    error
   \end{tabular}
  &Repeats/skips
  \\ \hline
  Count&$21$&$1$&$21$&$33$&$63$
  \end{tabular}
 }
\label{tab:realperf_statistics}
\end{table}
\begin{table}[t]
 \centering
 \caption{Statistics of differences in score times before and after repeats/skips in the performance
 data. ``Qu.'' is an abbreviation for quartile.}
  {\tabcolsep=4pt
 \begin{tabular}[t]{c|cccccc}
  Score time& Min. & 1st Qu. & Median & Mean & 3rd Qu. & Max. \\ \hline
  In second
  &$-84.83$&$-15.5$&$-7.75$&$-8.775$&$-1.875$&$45.750$\\
  In event
  &$-331$&$-44$&$-25$&$-23.35$&$-4$&$178$\\
 \end{tabular}
 }
\label{tab:realperf_repeatskip_statistics}
\end{table}

We collected $16$ audio recordings of clarinet performances
with a time range of $31$ to $213$ s (totally $28$ min $48$ s).
We requested an amateur clarinetist to freely practice
seven music pieces containing classical and popular music pieces
and nursery rhymes, partially from RWC music database
\cite{Goto2004}.
His performances were recorded with a vibration microphone
attached to the clarinet.

The performances were aligned to the notes in the scores by
one of the authors.
The total number of performed notes was $2672$, and 
Tab.~\ref{tab:realperf_statistics} lists
the count of errors and repeats/skips.
Tab.~\ref{tab:realperf_repeatskip_statistics} summarizes
differences in score times before and after repeats/skips
in the performance data,
and we see that they contain repeats/skips between remote score
positions.
Here, only breaks and pauses between notes longer than $500$ ms
were counted since it is difficult to accurately annotate offsets of
performed notes and short silent breaks and pauses between notes.
All transitions with $j\notin\nbh(i)$ were counted as repeats/skips,
where $i$ and $j$ denote stop and resumption positions.

\subsubsection{Results}
\label{sec:scofo_perf_result}
The parameters were same as in Sec.~\ref{sec:expcond_Bach10}.
To measure how well the algorithms followed repeats/skips,
we calculated a detection rate of repeats/skips
and the time interval between a repeat/skip and its detection, which we
call following time.
A repeat/skip was defined to be detected if there was a correctly
estimated frame until the next repeat/skip or the end of the audio
recording.

\begin{figure}[t]
 \centering
 \includegraphics[width=\columnwidth,clip]{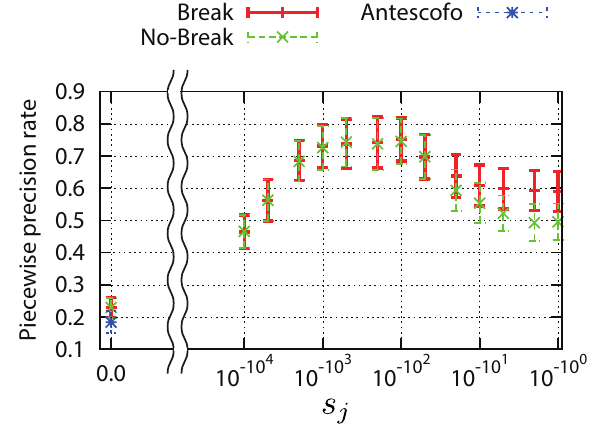}
 \caption{Average piecewise precision rates with standard errors with
 respect to $s_j$.
 The algorithms are same as in Fig.~\ref{fig:evalppr_Bach10}.}
 \label{fig:evalppr}
\end{figure}

\begin{table}[t]
 \centering
 \caption{Detection rates of repeats/skips for varying $s_{j}$.
 The algorithms are same as in Fig.~\ref{fig:evalppr_Bach10}.}
 \begin{tabular}[t]{c|ccc}
  $s_{j}$&Break&No-Break&Antescofo\\ \hline
  $1.0\times10^{-1}$&$58/63$&$60/63$&-\\
  $1.0\times10^{-5}$&$59/63$&$59/63$&-\\
  $1.0\times10^{-10}$&$59/63$&$60/63$&-\\
  $1.0\times10^{-50}$&$58/63$&$59/63$&-\\
  $1.0\times10^{-100}$&$56/63$&$57/63$&-\\
  $1.0\times10^{-500}$&$55/63$&$55/63$&-\\
  $1.0\times10^{-1000}$&$56/63$&$55/63$&-\\
  $1.0\times10^{-5000}$&$43/63$&$43/63$&-\\
  $0.0$&$13/63$&$13/63$&$4/63$\\
 \end{tabular}
\label{tab:evalFT_sum}
\end{table}

\begin{figure}[t]
 \centering
 \includegraphics[width=\columnwidth,clip]{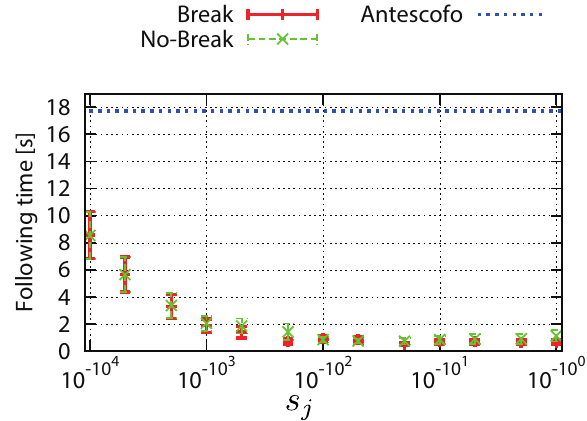}
 \caption{Average following time and standard error for varying $s_j$.
 The algorithms are same as in Fig.~\ref{fig:evalppr_Bach10}.
 For Antescofo, only the average following time is shown.}
 \label{fig:evalFT_meansd}
\end{figure}

Fig.~\ref{fig:evalppr} illustrates average PPRs with standard errors for
$\Delta=300$ ms.
Both proposed algorithms outperformed Antescofo at all $s_{i}$s,
clearly showing that the proposed algorithms are effective in
following performances with errors and repeats/skips.
A similar tendency was observed in PPR with $\Delta=100,500$ and $2000$ ms.
We also measured the effect of adding the pause states in
the proposed algorithms with $s_{i}=1.0\times10^{-100}$, and found
that it increased PPRs by $0.05$ on average.

Tab.~\ref{tab:evalFT_sum} summarizes the detection rates of
repeats/skips, and Fig.~\ref{fig:evalFT_meansd}
illustrates averages of following times over all detected
repeats/skips (average following times) and standard errors in
second.
Since the standard error for Antescofo was too large to display in the
figure, only the average value is shown.
Both proposed algorithms clearly outperformed Antescofo in the detection
rate and the following time.
For example, compared to Antescofo,
both proposed algorithms with $s_{j}=1.0\times10^{-100}$
detected $14$ times more repeats/skips and caught up with them
$20$ times faster in second.
These results show that the proposed models are effective for
repeats/skips.

The {\em break algorithm} (the {\em no-break algorithm}) with
$s_{j}=1.0\times10^{-100}$ detected $56$ ($57$) repeats/skips,
but failed to detect seven (six, respectively) repeats/skips.
These failures were caused by
the existence of sections and phrases similar to each other in the
scores (e.g. choruses in popular music)
and considerably short performances between repeats/skips.
For example, nine performances between repeats/skips were below five
seconds.

Most of the repeats/skips accompanied silent breaks,
but the {\em break algorithm} provided similar results
to the {\em no-break algorithm}.
This is because the top states associated with rests can play the same
role of the break state since these top states were connected
to all top states.

\begin{table}[t]
\centering
 \caption{Detection rates of repeats/skips for
 violin, clarinet, saxophone and bassoon data
 with simulated errors and repeats/skips.
 The algorithms are same as in Fig.~\ref{fig:evalppr_Bach10},
 and $s_j=1.0\times 10^{-1000}$ was used in both proposed algorithms.
 }
 \begin{tabular}{c|ccc}
  Musical instrument & Break & No-Break & Antescofo\\ \hline
  Violin & $13/13$ & $13/13$ & $2/13$ \\
  Clarinet & $11/11$ & $10/11$ & $4/11$ \\
  Saxophone & $11/12$ & $11/12$ & $2/12$ \\
  Bassoon & $10/10$ & $10/10$ & $0/10$
 \end{tabular}
 \label{tab:Bach10_detection_rate}
\end{table}

\begin{figure*}[t]
 \centering
 \subfigure[Average piecewise precision rates with standard errors]{
 \includegraphics[width=\columnwidth]{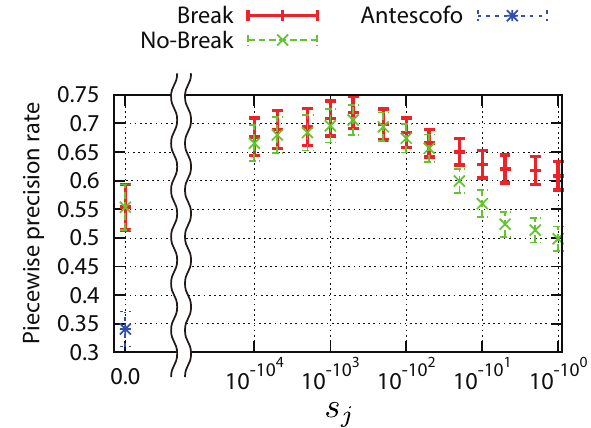}
 }
 \subfigure[Average following times with standard errors]{
 \includegraphics[width=\columnwidth]{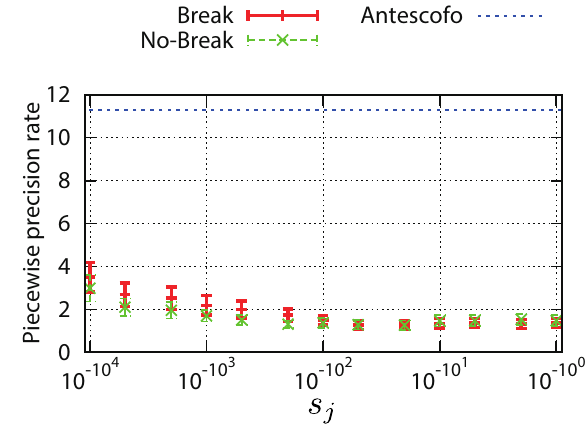}
 }
 \caption{(a) Average piecewise precision rates and (b) average
 following times with respect to $s_{j}$ for audio performances with
 simulated errors and repeats/skips.
 The algorithms are same as in Fig.~\ref{fig:evalppr_Bach10}, and
 only the average following time is shown for Antescofo in the right panel.}
\label{fig:exp_Bach10_simulated_repeats}
\end{figure*}

Furthermore, we measured following times and detection rates for
performances played by other musical instruments.
The audio recordings in the Bach10 dataset did not contain repeats/skips,
and we synthesized performances containing repeats/skips by randomly
jumped between breaks in each recording with a
probability of $0.1$ and inserting silent breaks at repeats/skips.
The durations of the breaks were sampled uniformly from $0.5$ to $30$
seconds and each synthesized performance was forced to contain at least
one repeat/skip.
After the synthesis, errors were simulated in the same way as in
Sec.~\ref{sec:error_simulate_Bach10}.
Tab.~\ref{tab:Bach10_detection_rate}
summarizes detection rates of repeats/skips for every musical
instrument.
The proposed algorithms with $s_{j}=1.0\times 10^{-1000}$ outperformed
Antescofo in the detection rate, and we found similar tendency in
the PPR and the following time as shown in
Fig.~\ref{fig:exp_Bach10_simulated_repeats}~(a) and (b),
respectively.
These results show that the proposed algorithms are also effective
in following performances with errors and repeats/skips
for various musical instruments.

A demonstration video of an automatic accompaniment system using the
{\em break algorithm} without the pause states
is available at \url{https://www.youtube.com/watch?v=fW6VKiC4k34} on
Youtube \cite{Sagayama2014}.
In the video, the {\em break algorithm} successfully
follows the performances during practice
and catches up the performances after repeats/skips within a few seconds.

\section{Discussions}
\label{sec:Discussions}
\subsection{Improvement of the Proposed algorithms}
\label{sec:improvement}
We now discuss possible extensions of the proposed algorithms.
The stop and resumption positions are not completely random,
and their distributions have certain tendencies in actual performances
\cite{ENakamura2014}.
For example, performers frequently resume from the first beats of bars
and the beginning of phrases, which reflects
performers' understanding of musical structures.
These tendencies can be incorporated in $s_{j},r_{i}$ in our performance
HMMs, and the accuracy and following times of the proposed algorithms
would improve \cite{ENakamura2014}.

Another method to improve the proposed algorithms
is to refine the model of the durations of performed events.
For this purpose, we can assign multiple bottom states to model the
duration \cite{Cano1999,Raphael1999,Orio2001}
or explicitly introduce its probability distribution \cite{Cont2010}.
This refinement is compatible with the proposed methods to reduce the
computational cost since they can be used regardless of the topology of
the bottom HMMs.

The proposed algorithms successfully followed clarinet performances
against tempo changes in the experiment and the demonstration video
in Sec.~\ref{sec:scofo_perf}.
However, the accuracy may deteriorate for the performances with large
tempo changes.
To suppress the deterioration, it would be effective to adequately change
$d_{i}$ on the fly, referring to estimated tempos.

\subsection{Extension to Polyphonic Music}
\label{sec:polyphony}
Although we have confined ourselves to monophonic performances,
let us briefly discuss the polyphonic case.
We can construct a performance HMM for polyphonic scores
similarly to the monophonic case.
By associating top states with musical events
(chords, notes and rests) in a polyphonic score,
the top HMM can be used without any change,
and insertions and deletions of chords, pauses between chords
and repeats/skips can be incorporated in the same way.
Importantly, the present methods to reduce the computational
complexity can be applied to the polyphonic case
since it is independent of details of the bottom HMMs.
On the other hand, we need to extend the bottom HMMs
to include chords.
Especially, errors may occur at every note in a chord,
and there are a combinatorially large number of possible forms of
errors for a large chord.
Although we could prepare spectral templates for all possible forms of
played chords and use a mixture distribution similarly to
Eq.~\eqref{eq:def_emsprob} in principle, it requires large computational
cost in estimating score positions.
However, the influence of note-wise errors in spectral differences is
generally less significant for a large chord,
and a bold approximation of neglecting note-wise errors would work
relatively well for such a case, which can serve as a practical method
to avoid the large computational cost.

There are other issues for polyphonic performances.
For example, notes in a chord are indicated to be performed
simultaneously in the score, but they can be actually performed
at different times.
Also, relative energy of notes in a chord depends on the performer.
Their treatment requires additional discussions and experiments, and the
extension to polyphonic performances is now under investigation.

\section{Conclusion}
\label{sec:conclusion}
We discussed score following of monophonic music performances with
errors and arbitrary repeats/skips by constructing a stochastic model of
music performance.
We incorporated possible errors in audio performances
into the model.
In order to solve the problem of large computational cost for following
arbitrary repeats/skips,
we presented two HMMs that describe a probability of repeats/skips with
a probability of stop positions
and a probability of resumption positions,
and derived computationally efficient algorithms.
We demonstrated real-time working of the algorithms with
scores of practical length (${\cal O}(10^3)$ to ${\cal O}(10^4)$
events).
Experimental evaluations using clarinet performance data showed that the algorithms outperformed
Antescofo in the accuracy of score following and the tracking ability of
repeats/skips.
In addition, we briefly discussed methods to improve the proposed
algorithms and extend them for polyphonic inputs.

\section*{Acknowledgements}
We thank Yuu Mizuno and Kosuke Suzuki for participating in the early
stage of this work, Naoya Ito for playing the clarinet, and Hirokazu
Kameoka for useful discussions.
This research was supported in part by
JSPS Research Fellowships for Young Scientists No.~15J0992 (T.~N.),
and JSPS Grant-in-Aid No.~15K16054 (E.~N.) and No.~26240025 (S.~S.).

\bibliographystyle{ieeetr}
\bibliography{simpstring20140909,refs_at_master,mybiblist}

\appendix
\subsection{List of important parameters} \label{sec:parameters}
Important parameters of the proposed models are listed in
Tab.~\ref{tab:list}.
  
\begin{table}[t]
 \centering
 \caption{Important parameters and their meanings of the proposed models.}
 \begin{tabular}{c|c}
   \begin{tabular}{c}
    Mathematical\\
    notation
   \end{tabular}
  & Meaning\\ \hline \hline
  $a_{i,j}$ & Transition probability of the top HMM\\
  $\pi_{i}$ & Initial probability of the top HMM\\ \hline
  $a_{l,l'}^{(i)}$ & Transition probability of the $i$-th bottom HMM\\
  $\pi_{l}^{(i)}$ & Initial probability of the $i$-th bottom HMM\\
  $e_{l}^{(i)}$ & Exiting probability of the $i$-th bottom HMM\\
  $b^{(i)}_{l}(\mathbf{y}_t)$ & Emission probability of state $(i,l)$
      for observation $\mathbf{y}_{t}$\\ \hline
  $\tilde{a}_{(i,l),(j,l')}$ &
      \begin{tabular}{c}
       Transition probability of the standard HMM\\
       obtained by flatting the two-level HMM
      \end{tabular}
       \\
  $\tilde{\pi}_{(i,l)}$ &
	\begin{tabular}{c}
	 Initial probability of the standard HMM\\
	 obtained by flatting the two-level HMM
	 \end{tabular}
	 \\
  $\tilde{b}_{(i,l)}(\mathbf{y}_t)$ &
      	\begin{tabular}{c}
      Initial probability of the states of the standard HMM\\
	 obtained by flatting the two-level HMM
	 \end{tabular}
	 \\ \hline
  $s_{j}$ &
      \begin{tabular}{c}
      The probability of stopping at event $j$\\
       before a repeat/skip
       \end{tabular}
  \\
  $r_{i}$ &\begin{tabular}{c}The probability of resuming a performance
	    at event $i$\\ after a repeat/skip \end{tabular}
 \end{tabular}
 \label{tab:list}
\end{table}

\subsection{Derivation of the No-Break Algorithm for $L>1$}
\label{sec:twolayeredHMM_outerproduct}
We now derive an efficient algorithm of computing $\alpha_{t,(i,l)}$ for
the performance HMM without the break state in the case of $L>1$.
Assuming that the transition probability of repeats/skips is described as 
a product of $s_{j}$ and $r_{i}$,
the transition probability of the standard HMM
$\tilde{a}_{(j,l'),(i,l)}$ for $j\notin \nbh(i)$ can be written as
\begin{align}
 \tilde{a}_{(j,l'),(i,l)}=e_{l'}^{(j)}s_{j}r_{i}\pi_{l}^{(i)},
\end{align}
and Eq.~\eqref{eq:ForwardAlgorithm_layered} for $t\geq1$ is
rewritten as
\begin{align}
\alpha_{t,(i,l)}
=&
\displaystyle
\tilde{b}_{(i,l)}(\mathbf{y}_t)
 \Big(
 \sum_{\substack{j\in\nbh(i)\\ l'=0,\cdots,L-1}}
 \alpha_{t-1,(i,l')}\tilde{a}_{(j,l'),(i,l)}
 \notag
\\
\displaystyle
&+
  r_{i}\pi_{l}^{(i)}
 \sum_{\substack{j\notin \nbh(i) \\ l'=0,\cdots,L-1}}
 \alpha_{t-1,(j,l')}e_{l'}^{(j)}s_{j}
 \Big).
 \label{eq:OuterProductHMM_Forward_layered_sum}
\end{align}
The first summation in the parentheses of 
Eq.~\eqref{eq:OuterProductHMM_Forward_layered_sum} is of ${\cal O}(L)$.
The second summation can be converted into 
\begin{align}
  &\sum_{\substack{j\notin \nbh(i) \\ l'=0,\cdots,L-1}}
 \alpha_{t-1,(j,l')}e_{l'}^{(j)}s_{j}
 \notag \\
 =&
 \sum_{\substack{j=0,\cdots,N-1\\ l'=0,\cdots,L-1}}
 \alpha_{t-1,(j,l')}e_{l'}^{(j)}s_{j}
 -
 \sum_{\substack{j\in\nbh(i)\\ l'=0,\cdots,L-1}}
 \alpha_{t-1,(j,l')}e_{l'}^{(j)}s_{j}.
 \label{eq:OuterProductHMM_Forward_layered_sum2}
\end{align}
The first summation of the right-hand side of 
Eq.~\eqref{eq:OuterProductHMM_Forward_layered_sum2} is independent of $i$
and thus it is sufficient to compute it once at each time step.
Hence, the total computational complexity at each time step is of ${\cal O}(LN)$.

\subsection{Derivation of the Break Algorithm for $L>1$}
\label{sec:twolayeredHMM_proposed}
Let us consider the performance HMM with the break state and with $L$
bottom states in each top state.
In the same way as Sec.~\ref{sec:break}, silent breaks at repeats/skips
can be introduced as top state $N$ (the break state) and arbitrary
repeats/skips are described with two-step transitions via the break state.
Since the transition probability of the standard HMM
$\tilde{a}_{(j,l'),(i,l)}$ is zero unless $j\in\nbh(i)\cup\{N\}$,
Eq.~\eqref{eq:ForwardAlgorithm_layered} for $t\geq 1$ and $i\neq N$ can
be rewritten as
\begin{align}
 \alpha_{t,(i,l)}=&
\tilde{b}_{(i,l)}(\mathbf{y}_{t})
\Big(
\sum_{\substack{j\in \nbh(i)\\ l'=0,\cdots,L-1}}
\alpha_{t-1,(j,l')}
\tilde{a}_{(j,l'),(i,l)}
& 
\notag
\\
&
\displaystyle
 +
 \sum_{l'=0}^{L-1}
\alpha_{t-1,(N,l')}\tilde{a}_{(N,l'),(i,l)}
\Big),
\label{eq:ForwardAlgorithm_wsilent_layered}
\end{align}
The second term in the parentheses of
Eq.~\eqref{eq:ForwardAlgorithm_wsilent_layered} for
each $i\neq N$ is of a constant computational complexity.
On the other hand, Eq.~\eqref{eq:ForwardAlgorithm_layered} for $t\geq 1$
and $i=N$ is converted into
\begin{align}
 \alpha_{t,(N,l)}=&
 \tilde{b}_{(N,l)}(\mathbf{y}_{t})
 \sum_{\substack{j=0,\cdots,N-1 \\ l'=0,\cdots,L-1}}
 \alpha_{t-1,(N,l)}\tilde{a}_{(j,l'),(N,l)}.
\end{align}
This computation is of ${\cal O}(LN)$
and hence the total computational complexity
is of ${\cal O}(LN)$ at each time step.

\newpage
\begin{IEEEbiography}[{\includegraphics[width=1in,height=1.25in,clip,keepaspectratio]{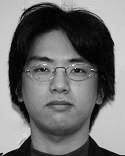}}]{Tomohiko Nakamura}
He received his B.E and M.S. degrees from the University of Tokyo,
Japan, in 2011 and 2013, respectively.
He is currently a Ph.~D. student at the University of Tokyo and a
 research fellow of Japan Society for the Promotion of Science (JSPS).
His research interests involve audio signal processing and statistical
 machine learning.
He received International Award from the Society of Instrument and
 Control Engineers (SICE) Annual Conference 2011,
SICE Best Paper Award (Takeda Award) in 2015,
and Yamashita SIG Research Award from the Information Processing Society
 of Japan (IPSJ) in 2015.
\end{IEEEbiography}

\begin{IEEEbiographynophoto}{Eita Nakamura}
He received a Ph.~D. in physics from the University of Tokyo in
2012. After having been a post-doctoral researcher at the National
Institute of Informatics and Meiji University, he is currently a
post-doctoral researcher at the Speech and Audio Processing Group at
Kyoto University. His research interests include music information
processing and statistical machine learning.
\end{IEEEbiographynophoto}

\begin{IEEEbiography}[{\includegraphics[width=1in,height=1.25in,clip,keepaspectratio]{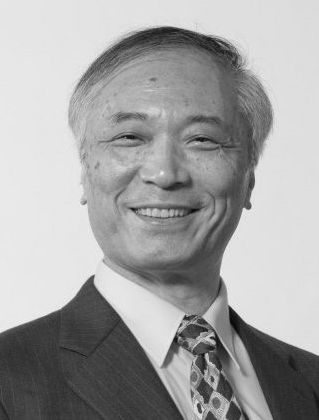}}]{Shigeki Sagayama}
He received the B.E., M.S., and
Ph.D. degrees from the University of Tokyo, Tokyo,
Japan, in 1972, 1974, and 1998, respectively, all in
mathematical engineering and information physics.
He joined Nippon Telegraph and Telephone Public
Corporation (currently, NTT) in 1974 and started his
career in speech analysis, synthesis, and recognition
at NTT Labs in Musashino, Japan. From 1990, he
was Head of the Speech Processing Department,
ATR Interpreting Telephony Laboratories, Kyoto,
Japan where he was in charge of an automatic
speech translation project. In 1993, he was responsible for speech recognition,
synthesis, and dialog systems at NTT Human Interface Laboratories,
Yokosuka, Japan. In 1998, he became a Professor of the Graduate School of
Information Science, Japan Advanced Institute of Science and Technology
(JAIST), Ishikawa. In 2000, he was appointed Professor at the Graduate
School of Information Science and Technology (formerly, Graduate School of
Engineering), the University of Tokyo. After his retirement from the University
of Tokyo, he is a Professor of Meiji University from 2014.
His major research interests include the processing
and recognition of speech, music, acoustic signals, handwriting, and images.
He was the leader of anthropomorphic spoken dialog agent project (Galatea
Project) from 2000 to 2003. Prof. Sagayama received the National Invention
Award from the Institute of Invention of Japan in 1991, the Director General's
Award for Research Achievement from the Science and Technology Agency
of Japan in 1996, and other academic awards including Paper Awards from
the Institute of Electronics, Information and Communications Engineers, Japan
(IEICEJ) in 1996 and from the Information Processing Society of Japan (IPSJ)
in 1995. He is a member of the Acoustical Society of Japan, IEICEJ, and IPSJ.
\end{IEEEbiography}

\end{document}